\documentstyle[twocolumn,eqsecnum,aps,prb,epsfig]{revtex}

\newcommand{\ybco}{$\mathrm{YBa_{2}Cu_{3}O_{7-\delta}}$}

\newcommand{\tk}{$T_{KT}$}

\newcommand{\loglog}{$\log(V)$ vs. $\log(I)$}
\newcommand{\ktb}{KTB}
\newcommand{\htcing}{high temperature superconducting}

\newcommand{\ffh}{FFH}

\newcommand{\jj}{Josephson junction}

\begin{document}

\draft
\author{D. R. Strachan~\cite{strachan} and
    C. J. Lobb}
\address{Center for Superconductivity Research, University of Maryland,
College Park, MD 20742}
\author{R. S. Newrock}
\address{Department of Physics, University of Cincinnati,
Cincinnati, OH 45421}
\title{Dynamic Scaling and
Two-Dimensional High-$\mathrm{\bf{T_c}}$ Superconductors}
\date{\today}

\twocolumn[\hsize\textwidth\columnwidth\hsize\csname@twocolumnfalse\endcsname

\maketitle

\begin{abstract}
There has been ongoing debate over the critical behavior of
two-dimensional superconductors; in particular for high $T_{c}$
superconductors.  The conventional view is that a
Kosterlitz-Thouless-Berezinskii transition occurs as long as
finite size effects do not obscure the transition.  However, there
have been recent suggestions that a different transition actually
occurs which incorporates aspects of both the dynamic scaling
theory of Fisher, Fisher, and Huse and the
Kosterlitz-Thouless-Berezinskii transition.  Of general interest
is that this modified transition apparently has a universal
dynamic critical exponent.  Some have countered that this apparent
universal behavior is rooted in a newly proposed finite-size
scaling theory; one that also incorporates scaling and
conventional two-dimensional theory.  To investigate these issues
we study DC voltage versus current data of a $\mathrm{12 \AA}$
thick \ybco\ film.  We find that the newly proposed scaling
theories have intrinsic flexibility that is relevant to the
analysis of the experiments.  In particular, the data scale
according to the modified transition for arbitrarily defined
critical temperatures between 0 K and 19.5 K, and the temperature
range of a successful scaling collapse is related directly to the
sensitivity of the measurement.  This implies that the apparent
universal exponent is due to the intrinsic flexibility rather than
some real physical property.  To address this intrinsic
flexibility, we propose a criterion which would give conclusive
evidence for phase transitions in two-dimensional superconductors.
We conclude by reviewing results to see if our criterion is
satisfied.

\end{abstract}



]





\section{Introduction}      \label{sec:intro}

There have been many reports of a two-dimensional (2D)
Kosterlitz-Thouless-Berezinskii (KTB) phase
transition~\cite{kosterlitz73_1jpc,kosterlitz74_1jpc,berezinskii71_1jetp}
in the cuprate
superconductors.~\cite{norton93_1prb,yeh89_1prb,vadlamannati91_1prb,stamp88_1prb,ying90_1prb,kim89_1prb,park00_1prb,matsuda92_1prl,walkenhorst92_1prl,ammirata99_1pc,ban89_1prb,teshima92_1pc,pierson99_1prb,fiory88_1prl,martin89_1prl}
Repaci~\emph{et al.}~\cite{repaci96_1prb} studied the ``most two
dimensional'' \htcing\ sample possible, a single unit cell thick
film of \ybco . They concluded that there was no phase transition
because a population of unbound vortices exists in this system
well below the temperature where a conventional \ktb\ transition
would occur. This conclusion was shown to be in agreement with
theory because the requirement that the sample be much smaller
than the perpendicular penetration depth\cite{beasley79_1prl} was
not met.~\cite{repaci96_1prb}

In contradiction, Pierson \emph{et al.}~\cite{pierson99_1prb} and
Ammirata \emph{et al.}~\cite{ammirata99_1pc} re-analyzed Repaci
\emph{et al.}'s data (along with those from others) by using the
general scaling ideas of Fisher, Fisher, and Huse
(FFH)~\cite{fisher91_1prb} and concluded that finite-size effects
do not obscure the KTB transition. They further concluded that
important details of the \ktb\ dynamics were not included in its
original formulation for superconductors by Halperin and
Nelson.~\cite{halperin79_1jltp} In this paper we will refer to
this Pierson and Ammirata analysis as the modified KTB scaling
analysis.

Later theoretical work by Colonna-Romano \emph{et
al.},~\cite{colonna-romano00_1condmat} Medvedyeva \emph{et
al.},~\cite{medvedyeva00_1prb,medvedyeva01_1physicaC} and Holzer
\emph{et al.}~\cite{holzer01_1prb} suggests that the KTB
transition \emph{can} be obscured by finite-size effects. This
work, in distinction from the earlier scaling
analysis,~\cite{ammirata99_1pc,pierson99_1prb} seems to support
Repaci~\emph{et al.}'s original conclusions.  That being said, all
of these theoretical works do however find it intriguing that the
modified KTB scaling analysis produces the large value of $z
\approx 6$ for a variety of 2D systems.

In addition, Medvedyeva \emph{et
al.}~\cite{medvedyeva00_1prb,medvedyeva01_1physicaC} have proposed
a finite-size scaling form that is also based on FFH-scaling
applied to 2D superconductors.  They argue that it is this
finite-size scaling, as opposed to the modified KTB scaling, that
is the correct scaling form for the data of Repaci~\emph{et al.}
They further argue that under certain conditions their finite-size
scaling is of the same form as the modified KTB scaling analysis
of Pierson and Ammirata.  Since this would mean that the modified
KTB transition only appears to occur while their finite-size
scaling is the correct physical description, Medvedyeva \emph{et
al.} label this a ``ghost"~\cite{medvedyeva00_1prb} transition.

The question is: Does the reoccurring large value of $z \approx 6$
suggest a common origin; possibly resulting from a modified KTB
transition or perhaps from a ``ghost" transition based on
finite-size scaling features in the vicinity of a KTB transition?

The work we present here indicates that the results of $z \approx
6$ from the modified KTB scaling analysis are not linked to
fundamental aspects of a KTB transition. This is due to the fact
that an opposite concavity criterion, similar to one proposed for
3D transitions in magnetic fields by Strachan \emph{et
al.},~\cite{strachan01_prl} is not satisfied by the experimental
data. Instead, we argue that the agreement between the many
systems analyzed by Pierson \emph{et al.} and Ammirata \emph{et
al.} in Refs.\ \onlinecite{pierson99_1prb} and
\onlinecite{ammirata99_1pc} (having $z \approx 6$) stems from the
fact that a single experimental data set has a tremendous amount
of flexibility when analyzed through modified KTB scaling.

In our work, we have carefully re-analyzed Repaci's data.  We are
able to successfully collapse the experimental data to the
modified KTB scaling theory~\cite{ammirata99_1pc,pierson99_1prb}
using a wide range of critical temperatures and exponents;
something discussed in the theoretical work of Medvedyeva \emph{et
al.}~\cite{medvedyeva00_1prb} only above the transition but which
we investigate in depth on experimental measurements at all
temperatures. In fact, we find that this modified KTB scaling is
still achievable when the critical temperature is defined to lie
well outside of the temperature range of the measurements and for
$z$ values much larger than $6$. This indicates that a single data
set could be found to agree with any sufficiently large $z$ value,
which could easily explain the agreement amongst the various 2D
systems. We also find that the modified KTB scaling analysis is
further weakened by its strong dependence on the voltage
resolution limit of the experiment, as was theoretically
demonstrated by Holzer \emph{et al.}~\cite{holzer01_1prb} and in
much the same way that the conventional vortex glass scaling
is.~\cite{haan99_1prb}

In distinction from modified KTB scaling, we find that a ``ghost"
transition based on the finite-size scaling of Medvedyeva \emph{et
al.}~\cite{medvedyeva00_1prb,medvedyeva01_1physicaC} does not
contain the same flexibility in determining the $z$.  Despite this
possible success of dynamic-scaling applied to a 2D
superconductor, we find that the Medvedyeva \emph{et al.}
finite-size scaling theory entails its own flexibility and we
argue that the experimental data fall within this realm of
flexibility.

In a related issue, Medvedyeva \emph{et
al.}~\cite{medvedyeva00_1prb} argue that Repaci~\emph{et al.}'s
measurements show vestiges of a conventional KTB transition; one
that would have occurred in the film had finite-size effects not
obscured it. They determine this obscured KTB transition to be at
a temperature 10K higher than the modified transition temperature
of Pierson \emph{et al.}~\cite{pierson99_1prb} and Ammirata
\emph{et al.}~\cite{ammirata99_1pc}, a view that is in accord with
the original conclusions of Repaci~\emph{et
al.}~\cite{repaci96_1prb}

We address this issue in Sec.\ \ref{sec:discussion}, where we use
our opposite concavity criterion to motivate a criterion for
determining a universal jump in the super-electron density at a
conventional KTB transition. We demonstrate the need for our
criterion by showing how one might be led to determine that a KTB
transition exists, surprisingly, directly between the modified
$T_{KT}$ (of Pierson \emph{et al.}~\cite{pierson99_1prb} and
Ammirata \emph{et al.}~\cite{ammirata99_1pc}) and the
higher-temperature obscured $T_{KT}$ (of Medvedyeva \emph{et
al.}~\cite{medvedyeva00_1prb}).  We argue that this perplexing
predicament can be clarified by using our KTB concavity criterion.
We conclude by reviewing theoretical and experimental results to
see if they satisfy our KTB criterion. We find intriguing
agreement with only one experimental data
set~\cite{matsuda92_1prl} in the literature on a
high-$\mathrm{{T_c}}$ film and we compare this to the measurements
of Repaci~\emph{et al.}~\cite{repaci96_1prb}


\section{Scaling Applied to the Kosterlitz-Thouless-Berezinskii
transition}     \label{sec:theory}

The dynamic scaling approach of Pierson \emph{et
al.}~\cite{pierson99_1prb} and Ammirata \emph{et
al.}~\cite{ammirata99_1pc} is based on work by Fisher, Fisher, and
Huse (FFH).~\cite{fisher91_1prb} \ffh\ predict that the voltage,
$V$, across a superconductor with an applied current, $I$, at
temperature, $T$, should vary as

\begin{equation}
V=I \xi^{D-2-z} \chi_{\pm} \left( {I\xi^{D-1}}\over{T} \right),
\label{eq:scaling}
\end{equation}
where $\xi$ is the superconducting coherence length, $D$ is the
dimensionality, and $z$ is the dynamic exponent. The two
unspecified functions, $\chi_\pm$, apply above ($+$) and below
($-$) the transition temperature $T_c$.

Repaci's $\mathrm{12 \AA}$ thick film is two dimensional; for two
dimensions Eq.\ (\ref{eq:scaling}) becomes

\begin{equation}
V=I \xi^{-z} \chi_{\pm} \left( {I\xi}\over{T}
\right)= I \xi^{-z} \chi_{\pm} \left( x \right). \label{eq:2D_scaling}
\end{equation}
If a factor of $\left({I\xi}/T \right)^z$ is factored out of
$\chi_{\pm}$ and the remaining function is named
$\varepsilon_{\pm}$, we can rewrite Eq.\ (\ref{eq:2D_scaling}) as

\begin{equation}
{I \over T} \left( I \over V \right)^{1/z} =\varepsilon_{\pm}
(I\xi\ / T).  \label{eq:Pierson_scaling}
\end{equation}
Eq.\ (\ref{eq:Pierson_scaling}) is sometimes preferred for
analysis~\cite{ammirata99_1pc,pierson99_1prb} of data because
$\xi$, which is expected to diverge at $T_c$, is present only in
the argument to the scaling function, $\varepsilon_{\pm}$.

At $T_c$ the right-hand side of Eq.\ (\ref{eq:Pierson_scaling})
becomes $\varepsilon_{\pm} ( \infty )$ since $\xi$ diverges.  For
an applied $I$ the measured $V$ should be some non-zero finite
value at $T_c$, so that the left-hand side of Eq.\
(\ref{eq:Pierson_scaling}) is a non-zero finite value at $T_c$.
Therefore, $\varepsilon_{\pm} ( \infty )$ must also be non-zero
and finite at $T_c$.  Furthermore, for a non-zero applied current
the left-hand side of Eq.\ (\ref{eq:Pierson_scaling}) should be
continuous and smooth since non-analytic points, \emph{i.e.},
critical points, are only approached as $I \rightarrow 0$. This
requires that $\varepsilon_{+} ( \infty )=\varepsilon_{-} ( \infty
)=A$, where $A$ is a non-zero finite constant.  By setting $T=T_c$
and substituting $A$ in for the right-hand side of Eq.\
(\ref{eq:Pierson_scaling}) we can solve for the $I-V$ relation

\begin{equation}
V \propto I^{z+1} , \label{eq:power_law}
\end{equation}
valid at $T_c$.~\cite{standard_KTB_theory} It is important to note
that Eq.\ (\ref{eq:power_law}) is also valid at temperatures and
currents which make the argument, $x$, of $\varepsilon_{\pm}
\left( x \right)$ large since this causes Eq.\
(\ref{eq:Pierson_scaling}) to go to the same limiting form. Above
$T_{c}$, $\chi_+$ is expected to be constant in the
$x\rightarrow0$ limit of Eq.\ (\ref{eq:2D_scaling}). In this
limit, Eq.\ (\ref{eq:2D_scaling}) becomes

\begin{equation}
{V \over I} = R_{L} \propto \xi^{-z} \label{eq:R_L}.
\end{equation}

Having reviewed the well-known FFH scaling results for $D=2$, we
now discuss a proposed connection between them and the KTB
transition.

According to KTB theory one expects power law $I-V$ relations for
$T<T_{KT}$ of the form $V \sim I^{a(T)}$ due to current induced
unbinding of vortex pairs.  This occurs because vortex pairs with
size greater than $r_c$ and under the influence of an applied 2D
current density, $j_{2D}$, are repelled rather than attracted,
where (in SI units)

\begin{equation}
r_c= {{\Phi_{0}} \over { 2 \pi  ( {{\epsilon_{\infty}} \over
{\epsilon_{0}}}  ) \mu_0 j_{2D} \lambda_{\bot}}} .\label{rc}
\end{equation}
In the above relation, $\Phi_{0}$ is the flux quantum,
$\lambda_{\bot}$ is the bare (unrenormalized) penetration depth
for a thin film, $\mu_0$ is the permeability of vacuum,
$\epsilon_{0}$ is the permitivity of vacuum, and
$\epsilon_{\infty} \approx \epsilon(r_c)$ is the dielectric
constant which takes into account screening of pairs of size $r_c$
by other much smaller pairs.  As long as $r_c$ is large enough
such that $\epsilon(r_c)$ is approximately constant, \emph{i.e.},
$\epsilon_{\infty}$, one expects $V \sim I^{a(T)}$.

Harris \emph{et al.}~\cite{harris91_1prl} and, recently,  Pierson
\emph{et al.}~\cite{pierson99_1prb} and Ammirata \emph{et
al.}~\cite{ammirata99_1pc} have proposed using the KTB correlation
length,

\begin{equation}
\xi \propto e^{\sqrt{b / (\left| T-T_{KT} \right| /T_{KT})}},
\label{eq:xi_KT}
\end{equation}
in FFH's results with $T_{KT}$ replacing $T_c$.  This is a
surprising proposition because the KTB correlation length only has
the form of Eq.\ (\ref{eq:xi_KT}) above $T_{KT}$ $ \left( \xi_+
\propto e^{\sqrt{b / (\left| T-T_{KT} \right| /T_{KT})}} \right)$.
Below $T_{KT}$ the KTB correlation length, $\xi(T<T_{KT})$, is
infinite,~\cite{kosterlitz74_1jpc,colonna-romano00_1condmat,medvedyeva00_1prb}
which implies that the scaling relations (Eqs.\
(\ref{eq:scaling}-\ref{eq:Pierson_scaling})) should not be used
below $T_{KT}$.  (Note that $\xi(T<T_{KT})=\infty$ implies that
Eq.\ (\ref{eq:power_law}) holds for all $T \leq T_{KT}$.) This has
been countered through a careful renormalization group study by
Pierson and Valls,~\cite{pierson00_1prb} where they argue that a
physically relevant correlation length can be defined below
$T_{KT}$.  They define this physically motivated $\xi(T<T_{KT})$
as the point where the dielectric constant is essentially at its
asymptotic value, determined as the length scale yielding an
arbitrarily-chosen small fugacity.

However, our following derivation demonstrates the inconsistency
between FFH scaling and KTB theory regardless of the physical
interpretation of the correlation length below $T_{KT}$.  Our
argument is based on the fact that the $I-V$ relations below
$T_{KT}$ are of the form $V \sim I^{a(T)}$ as $I \rightarrow 0$
regardless of the physical interpretation of the correlation
length.

We demonstrate this inconsistency by taking $\left(
\partial logV/ \partial logI \right)_T$ of Eq.\
(\ref{eq:2D_scaling}) which yields~\cite{strachan02_proc}

\begin{equation}
{\left({{\partial logV} \over {\partial logI}} \right)}_T = 1 +
{{\partial log \chi_{\pm}(x)} \over {\partial x}}x=F_{\pm}(x).
\label{der1}
\end{equation}
This says that $\left( \partial logV/ \partial logI \right)_T$
should also scale as a function of ${I\xi/{T}}$.  In this regime
Eq.\ (\ref{der1}) becomes

\begin{equation}
a(T)= {\left({{\partial logV} \over {\partial logI}} \right)}_T =
F_{-} \left( {{I\xi} \over {T}} \right). \label{der2}
\end{equation}
The left-hand side of Eq.\ (\ref{der2}) depends only on
temperature whereas the right side depends on both $I$ and $T$
together as ${{I\xi} \over {T}}$.  This is impossible for
non-trivial choices of $F_{-}$ and thus FFH scaling and KTB theory
are inconsistent below $T_{KT}$.

Although Eq.\ (\ref{der2}) demonstrates inconsistency between Eq.\
(\ref{eq:scaling}) and KTB theory, we point out that making
$z=z(T)$, via Eq.\ (\ref{eq:power_law}), and
$\xi(T<T_{KT})=\infty$, scaling still applies below
$T_{KT}$.~\cite{medvedyeva00_1prb,jensen00_1prb}  This is the
premise of the finite-size scaling of Medvedyeva \emph{et
al.}~\cite{medvedyeva00_1prb}  One of the main points of that work
is the apparent existence of the scaling
form~\cite{medvedyeva00_1prb}

\begin{equation}
{{V} \over {IR(T)}} = h(IL {\mathrm{g}}_L(T)),
\label{minn_scaling}
\end{equation}
for smaller values of $IL {\mathrm{g}}_L(T)$.  In this relation,
$R(T)$ is the linear resistance of the low current ohmic tail at
temperature $T$, $L$ is the size of the 2D sample, and
${\mathrm{g}}_L(T)$ is an unspecified function of $T$ and $L$ that
permits a data collapse.

Having outlined the pertinent theory and shown the incompatibility
between KTB theory and FFF scaling through Eq.\ (\ref{der2}), we
will now demonstrate how the modified KTB scaling analysis may
lead one to believe that the two theories are compatible.


\section{The Modified KTB Scaling Analysis} \label{sec:data_conventional}

\begin{figure}
\epsfig{file=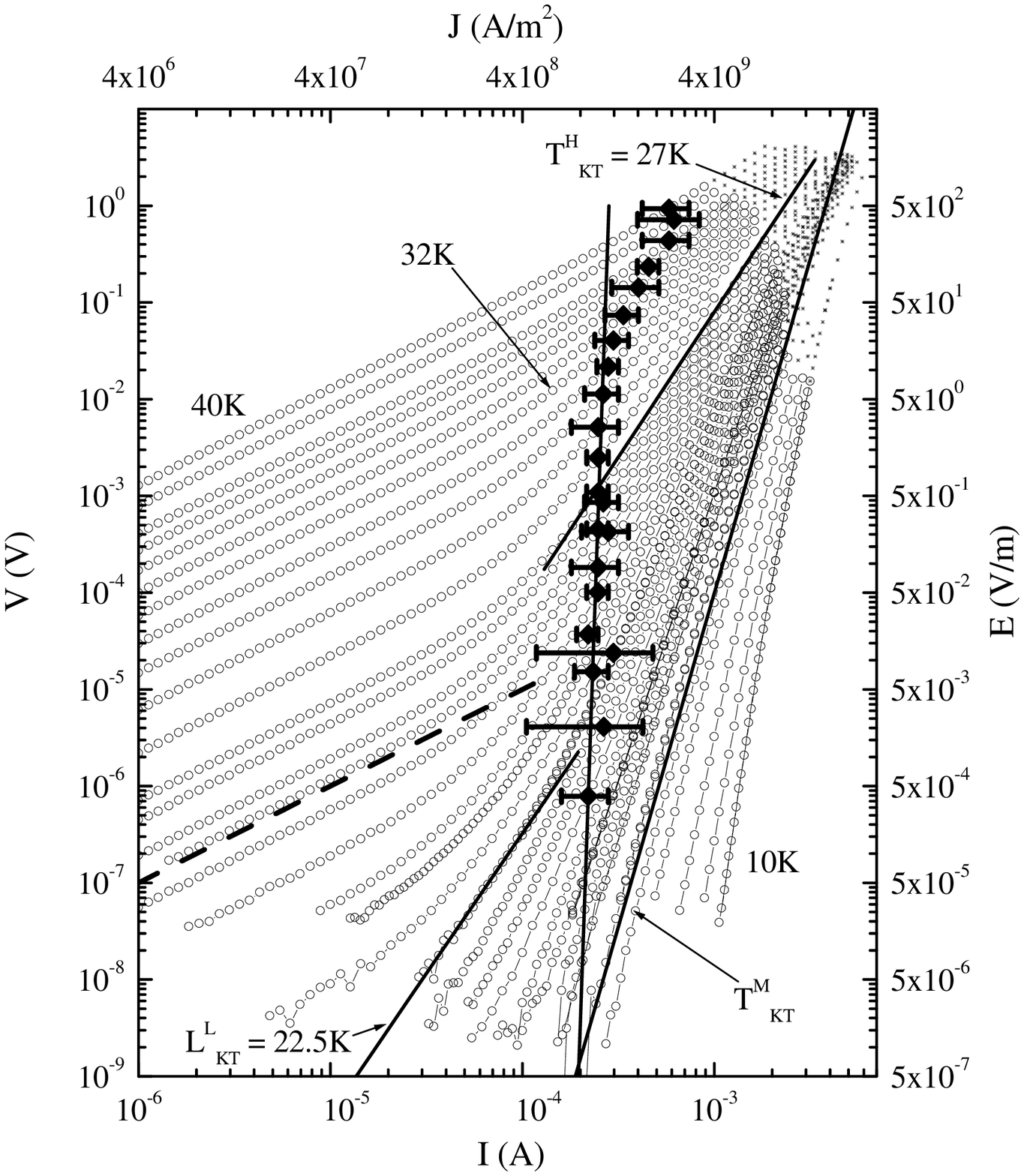,clip=,width=\linewidth} \caption{$I-V$
isotherms for a $12\mathrm{\AA}$ thick \protect\ybco\ film (from
Ref.\ \protect\onlinecite{repaci96_1prb}). The dashed line has a
slope of one, while the solid line labelled $T_{KT}^{M}$
represents the critical power law $I-V$ curve according to the
analysis of Refs.\ \protect\onlinecite{ammirata99_1pc} and
\protect\onlinecite{pierson99_1prb}. The solid diamonds are the
$I_{co}$ values where the isotherms change from non-ohmic to ohmic
behavior.} \label{fig:IV}
\end{figure}

Fig.\ \ref{fig:IV} shows a $\log - \log$ plot of Repaci \emph{et
al.}'s~\cite{repaci96_1prb} $I-V$ isotherms from a 12\AA \ thick
laser ablated \ybco\ film. Straight lines on this plot indicate
power law behavior, with the power equal to the slope of the line.
The dashed line on the left has a slope of 1, characteristic of
ohmic response.  Note that at low currents many isotherms have
ohmic tails.

Following the analysis of Pierson \emph{et
al.}~\cite{pierson99_1prb} and Ammirata \emph{et
al.}~\cite{ammirata99_1pc}, the solid line labelled
``$T_{KT}^{M}$" (Modified $T_{KT}$) in Fig.\ \ref{fig:IV} is a fit
to the isotherm which seems to separate those curves with ohmic
tails from those without. Pierson \emph{et al.} and Ammirata
\emph{et al.} define this isotherm to be the critical temperature,
$T_{KT}^{M} \approx 17.6\mathrm{K}$.

We note that $T_{KT}^{M}$ occurs when $V \propto I^{6.9}$ in the
data. This disagrees with the conventional KTB
analysis,~\cite{halperin79_1jltp} where the critical isotherm is
cubic, $V \propto I^{3}$, assuming that the length probed by the
applied current is large enough so that $\epsilon(r)$ is
completely renormalized. The only isotherm of Fig.\ \ref{fig:IV}
which is cubic at high currents is near 27K, denoted as
$T_{KT}^{H}$, which we demonstrate with the solid line fit (at
high currents) with $\mathrm{slope}=3$. However, this isotherm has
an ohmic tail at low currents, a signature of free vortices.
Medvedyeva \emph{et al.}~\cite{medvedyeva00_1prb} have recently
suggested that a conventional KTB transition would have occurred
near this $T_{KT}^{H}$ if finite-size effects had not obscured it;
a view in accord with the original arguments of Repaci \emph{et
al.}~\cite{repaci96_1prb}

Repaci \emph{et al.} originally argued~\cite{repaci96_1prb} that
the finite 2D penetration depth caused these free vortices below
$T_{KT}^{H}$. This is supported by Eq.\ (\ref{rc}), assuming that
the finite penetration depth becomes important once $r_c \approx
\lambda_{\bot}$. The ohmic tails of Fig.\ \ref{fig:IV} set in
roughly when $I \approx 1 \times 10^{-4} \mathrm{A}$, which
corresponds to $j_{2D}={I \over W} \approx {{1 \times 10^{-4}
\mathrm{A}} \over {2 \times 10^{-4} \mathrm{m}}}$, where $W$ is
the $200 \mu m$ width of the film.  A low density of vortex pairs
implies $\epsilon(r_c) \approx \epsilon_{\infty} \approx
\epsilon_{0}$, so that we can use Eq.\ (\ref{rc}) with
$\epsilon_{0}$ replacing $\epsilon_{\infty}$. Substituting
$\lambda_{\bot}$ for $r_c$ in Eq.\ (\ref{rc}), we find
$\lambda_{\bot} \approx 23 \mu m$, which is a reasonable (order of
magnitude) value for a unit cell thick YBCO film.


Pierson \emph{et al.}~\cite{pierson99_1prb} and Ammirata \emph{et
al.}~\cite{ammirata99_1pc} however, proposed that this same data
is well described by their modified KTB scaling theory.  Since
this modified analysis relaxes the constraint on the power of the
critical isotherm (\emph{i.e.} the $z$ exponent), their modified
\tk\ ($T_{KT}^{M}$) can occur at much lower temperatures. Thus,
Pierson \emph{et al.}~\cite{pierson99_1prb} and Ammirata \emph{et
al.}~\cite{ammirata99_1pc} argued that the ohmic tails below 27K
result from $T$ being above their modified $T_{KT}^{M}$, rather
than a finite-size effect.

A close look at the $I-V$ curves in Fig.\ \ref{fig:IV} reveals
that at high currents the isotherms near $T_{KT}^{M}$ bend away
from power law dependence towards ohmic behavior. Pierson \emph{et
al.}~\cite{pierson99_1prb} and Ammirata \emph{et
al.}~\cite{ammirata99_1pc} argued that the sample is driven normal
at these high currents and so they omit these data (denoted as
smaller symbols in Fig.\ \ref{fig:IV}) in their analysis. This
could be a reasonable assumption if the critical current density,
$J_c$,  for Repaci's film is assumed equal to the value for a
12\AA\ thick single crystal~\cite{tinkham_crit_cur} ($J_c \approx
10^{10} \mathrm{{{A} \over {m^2}}}$). This gives $I_c \approx 2
\times 10^{-3} \mathrm{A}$, which is approximately where the high
current down turn takes place. Following Pierson's and Ammirata's
prescription, we will also omit these data from the analysis of
this paper.

To find $b$ (Eq.\ (\ref{eq:xi_KT})) from theory one uses Eq.\
(\ref{eq:R_L}) in the form
\begin{equation}
\log_{10}({R_{L}})=\{ -z\sqrt{b} \log_{10}(e) \}
\sqrt{{T_{KT}^{M}} / {\left| {T-T_{KT}^{M}} \right|} } + C,
\label{eq:logRL}
\end{equation}
with $C$ a constant. $\mathrm{Log}_{10}(R_L)$ vs.
$\sqrt{{T_{KT}^{M}} / \left| {T-T_{KT}^{M}} \right| }$ with
$T_{KT}^{M}= \mathrm{17.6K}$ is plotted in the inset of Fig.\
\ref{fig:collapse_comparison}(a). According to modified KTB
scaling theory this should be linear, which it is, with slope
$-z\sqrt{b}\log_{10}(e)$.

Using the three parameters, $T_{KT}^{M}$, $b$, and $z$, Refs.\
\onlinecite{pierson99_1prb} and \onlinecite{ammirata99_1pc}
demonstrated that Repaci's data collapse to Eq.\
(\ref{eq:Pierson_scaling}).  [They also allowed for slight
variation of these parameters in order to optimize the scaling
collapse. This accounts for the slight deviation of the power-law
fit (using their $z$ exponent) from the data in Fig.\
\ref{fig:IV}.]

The scaling collapse is shown in Fig.\
\ref{fig:collapse_comparison}(a), where the higher current data is
towards the right and the lower temperatures are higher along the
vertical axis. All measurements taken below 17.6K fall on the
higher of the two curves ($\varepsilon_{-}$), while the rest are
described by $\varepsilon_{+}$.

There are two regions where the data seem to deviate slightly from
the scaling functions; these are indicated with arrows in Fig.\
\ref{fig:collapse_comparison}(a). The solid arrow points to data
which seem to bend away from the scaling function at high
currents. It is possible to argue that not enough of the high
current data (where, presumably, the sample is driven normal) have
been discarded from the analysis, as discussed above. The same may
be argued for the region indicated in Fig.\
\ref{fig:collapse_comparison}(a) by the dashed arrow. Thus, one
might conclude that there is good agreement with a modified KTB
scaling theory since the low current data seems to scale. This was
the conclusion in Refs.\ \onlinecite{pierson99_1prb} and
\onlinecite{ammirata99_1pc}.


\section{Further Analysis}  \label{sec:unconventional}

When the data of Fig.\ \ref{fig:IV} are examined closely, a
problem with the modified scaling analysis becomes apparent.
$T_{KT}^{M}$ occurs at the beginning of the region where the
voltmeter's sensitivity is no longer adequate to follow the $I-V$
curves down to currents sufficiently low for detecting the ohmic
tail. Since this ohmic tail is taken as evidence for being above
$T_{KT}^{M}$ in the modified KTB theory, its detection is crucial
in determining the critical temperature.

To quantify this, we determine the current, $I_{co}$, at which the
crossover from non-ohmic to ohmic behavior occurs as a function of
temperature.  $I_{co}$ is defined as the point where the slope is
a fraction, $f$, between the maximum slope of the isotherm and the
minimum. (The minimum is 1 for isotherms with an ohmic tail.) We
chose $f = 0.9$ although other choices, such as $f = 0.7$, do not
make significant changes (with $I_{co}$ lowering only by a factor
of about 1.7).

The solid diamonds in Fig.\ \ref{fig:IV} are these crossover
currents. The $I_{co}$ below 32K are roughly constant and slightly
greater than $10^{-4}\mathrm{A}$.  This is important because it is
the current where Pierson \emph{et al.}'s~\cite{pierson99_1prb}
and Ammirata \emph{et al.}'s~\cite{ammirata99_1pc} critical
isotherm of Fig.\ \ref{fig:IV} meets the $\mathrm{1nV}$ resolution
level of the voltmeter. A straight line through these points
intersects the critical isotherm of Refs.\
\onlinecite{pierson99_1prb} and \onlinecite{ammirata99_1pc}
\emph{at the resolution limit of the experiment.} (Repaci \emph{et
al.}~\cite{repaci96_1prb} realized this as well, which they
demonstrated through ${\left({{\partial logV} \over {\partial
logI}} \right)}_T$ plots in their Fig. 3.)

There are two possible explanations for this. The first is that
$T_{KT}^{M}$ is, by remarkable coincidence, precisely the
temperature where the voltmeter runs out of resolution, and the
$I-V$ curves below 17.6K do not have ohmic tails. The second is
that the curves below 17.6K have ohmic tails \emph{beyond the
voltmeter's resolution.}  The latter implies that there is no
phase transition at 17.6K, and the agreement between Repaci
\emph{et al.}'s data and the modified KTB transition is due to the
modified scaling analysis being much too flexible.

\subsection{Further analysis for $T_{KT}^{M} < 17.6 \mathrm{K}$}

To distinguish between these two possibilities, we re-analyze the
data using several lower values of modified $T_{KT}^{M}$. We find
the $z$ value at each $T_{KT}^{M}$ by setting the slope of the
steepest portion of that isotherm equal to $z + 1$ (Eq.\
(\ref{eq:power_law})).  Once $z$ is found for a modified
$T_{KT}^{M}$, we determine a value for $b$ exactly as is done in
Sec.\ \ref{sec:data_conventional}.  As with Pierson \emph{et
al.}'s~\cite{pierson99_1prb} and Ammirata \emph{et
al.}'s~\cite{ammirata99_1pc} analysis, we allow for a slight
change in $b$ and $z$ in order to optimize the data collapse.

\begin{figure}
\epsfig{file=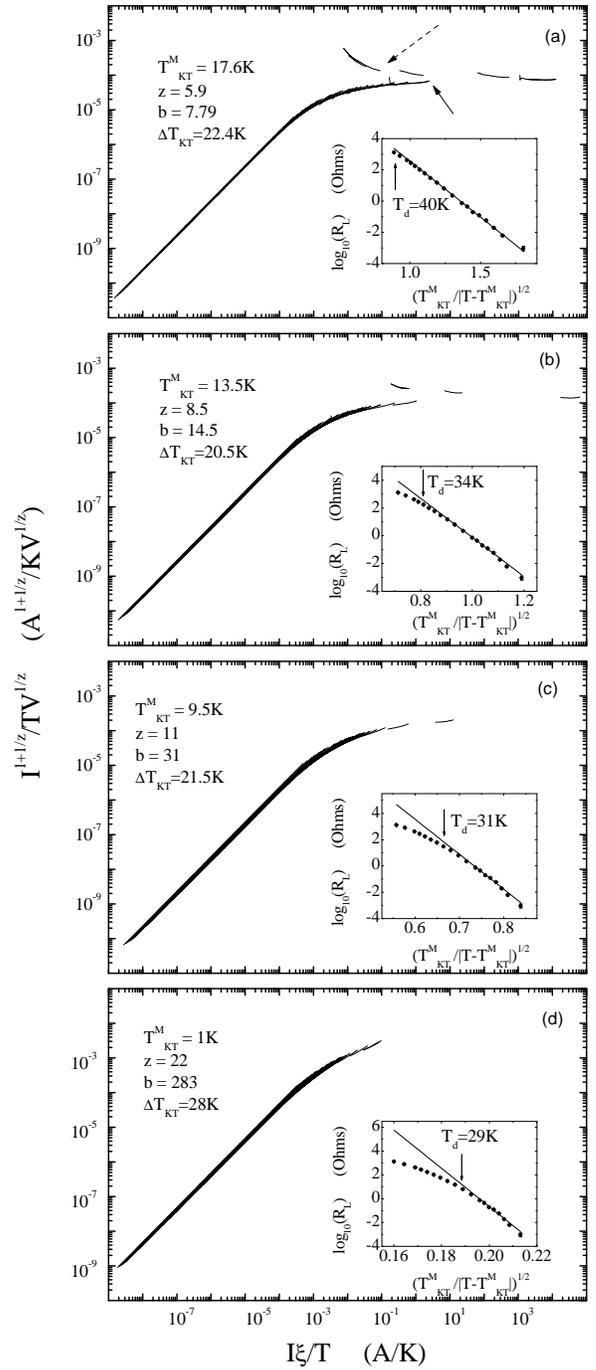,clip=,width=3.1in}
\caption{Scaled curves from
Ref.~\protect\onlinecite{repaci96_1prb} for various defined
$T_{KT}^{M}$ values from 17.6K down to 1K. The arrows in (a) point
out locations of the collapses where the high current data seems
to veer off the scaling functions.}
\label{fig:collapse_comparison}
\end{figure}

Starting with a $T_{KT}^{M}$ of 13.5K, we find $z=8.5$ and
$b=14.5$. For this $T_{KT}^{M}$ all the data that scales in Fig.\
\ref{fig:collapse_comparison}(a) also scales in Fig.\
\ref{fig:collapse_comparison}(b). The only difference in the
scaling collapse is that some data from $\varepsilon_{-}$ have
shifted to $\varepsilon_{+}$. As $T_{KT}^{M}$ is further lowered,
more data is shifted. When $T_{KT}^{M}$ reaches the low
temperature limit of the experiment, at about 10K (Fig.\
\ref{fig:collapse_comparison}(c)), there ceases to be anything
falling on the scaling function $\varepsilon_{-}$, which is
expected since there is no further data below this temperature.

We can lower $T_{KT}^{M}$ still further and still obtain a data
collapse. These scaling collapses are accompanied by much larger
$z$ and $b$ values, as is noted in Fig.\
\ref{fig:collapse_comparison}.


\subsection{Further analysis of ohmic tails} \label{RL_fit_section}

The inset of Fig.\ \ref{fig:collapse_comparison}(b) shows the
$R_{L}$ plot for $T_{KT}^{M}= \mathrm{13.5K}$. In this plot
$\log_{10}(R_L)$ deviates from linearity at about 33K, denoted
$T_d$. (Note that higher temperatures are located towards the
upper left.)  Since linear dependence of $\log_{10}(R_L)$ vs.
$(T_{KT}^{M}/(T-T_{KT}^{M}))^{1/2}$ is only expected near
$T_{KT}^{M}$ where critical fluctuations become important, this
does not imply a deviation from the modified KTB scaling theory.
In fact, the region of temperature ($\Delta
T_{KT}=T_{d}-T_{KT}^{M}$) where critical fluctuations are
important according to the $\log_{10}(R_L)$ plots in Fig.\
\ref{fig:collapse_comparison} is relatively independent of the
choice of $T_{KT}^{M}$.

We will denote the lowest temperature at which $R_L$ is measurable
as $T_{cut}$, which is 23K for Repaci's measurements. Although
Eq.\ (\ref{eq:logRL}) is non-analytic about $T_{KT}^{M}$, it
\emph{is} analytic at $T_{cut}>T_{KT}^{M}$.  Thus, we can Taylor
expand Eq.\ (\ref{eq:logRL}) about $T_{cut}$ as
\begin{eqnarray}\label{expand_RL}
\log_{10}({R_{L}}) & = & C - a_0(T_{cut}-T_{KT}^{M})^{-{1 \over
2}}
\\ & &  + {a_0 \over 2}(T_{cut}-T_{KT}^{M})^{-{3 \over 2}}(T-T_{cut}) \nonumber \\ & & -
{3a_0 \over 8}(T_{cut}-T_{KT}^{M})^{-{5 \over 2}}(T-T_{cut})^2 +
\cdots
\nonumber \\
a_0 & = & z\sqrt{bT_{KT}}\log_{10}(e). \label{a_0_for_expansion}
\end{eqnarray}
Given any arbitrary analytic function of temperature for the
linear $I-V$ behavior, $f_L(T)$, of the form
\begin{eqnarray}\label{arb_analytic}
\log_{10}(f_L(T)) = c_0+c_1(T-T_{cut}) & & \\
 & & +c_2(T-T_{cut})^2 + \cdots, \nonumber
\end{eqnarray}
this can be recast into the form of Eq.\ (\ref{expand_RL}), as
follows.

First we note that regardless of the values of $T_{KT}^{M}$, $z$,
$T_{cut}$, and $b$, we can always find a $C$ such that $c_0 = C -
a_0(T_{cut}-T_{KT})^{-{1 \over 2}}$. Likewise, for any
$T_{KT}^{M}$, z, and $T_{cut}$ we can always find a $b$ which
permits $c_1={a_0 \over 2}(T_{cut}-T_{KT})^{-{3 \over 2}}$. Since
this agreement will exist only close to $T_{KT}^{M}$, which is the
theoretical expectation, this test will always confirm KTB
scaling, modified or conventional.  (This is the flexibility in
this analysis implied in the discussion by
Mooij.~\cite{mooij94_1nato})

The other three parameters are nominally determined directly
through the measurements, as is the case for $T_{KT}^{M}$ and $z$,
or its limits, as  with $T_{cut}$.  However, these three
parameters have some flexibility, especially when selecting only
certain cuts of the data (as was done at high currents with
Pierson \emph{et al.}'s and Ammirata \emph{et al.}'s analysis of
Repaci's data) or when performing ``eye ball" data collapses where
$T_{KT}^{M}$ and $z$ can both be varied independently. This
flexibility allows for better apparent agreement with scaling
theory over an even larger temperature range.

Having demonstrated the large flexibility of the modified scaling
analysis for $T_{KT}^{M} < 17.6 \mathrm{K}$, we now examine higher
transition temperatures, $T_{KT}^{M} > 17.6 \mathrm{K}$.

\subsection{Further analysis for $T_{KT}^{M} > 17.6 \mathrm{K}$}

\begin{figure}
\epsfig{file=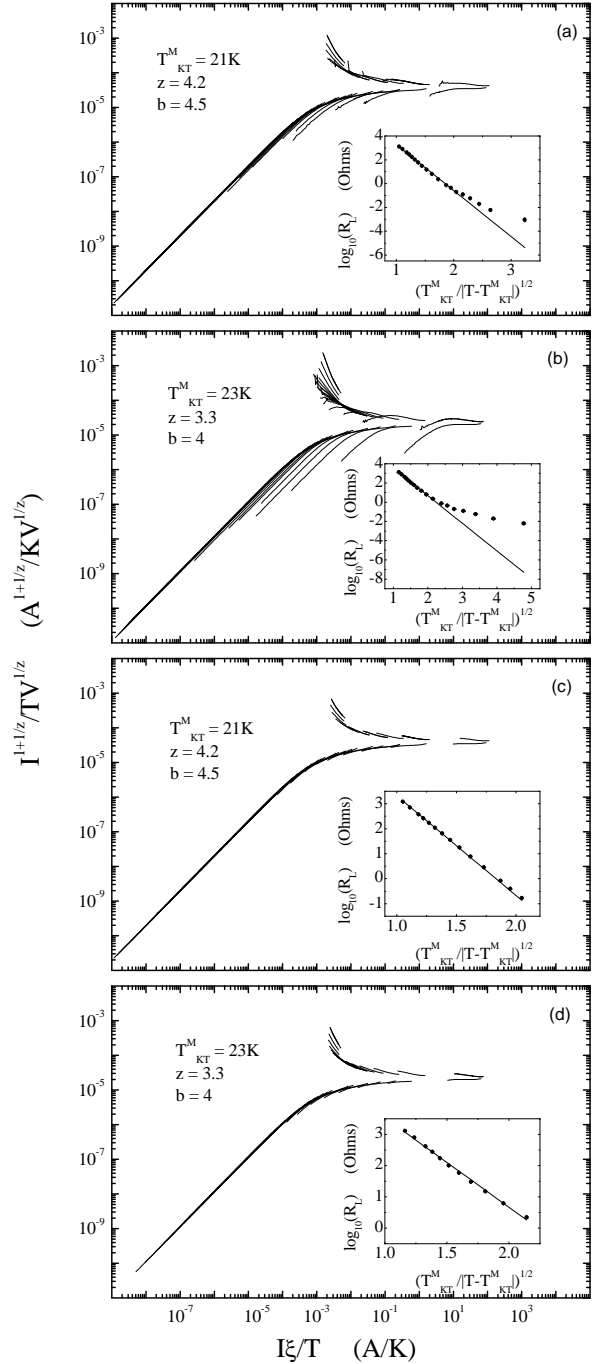,clip=,width=3.1in}
\caption{(a) and (b) show failed scaling collapses for
$T_{KT}^{M}$ equal to 21 and 24K respectively. (c) and (d) show
the same scaling attempts when resolution limits are imposed upon
the data. (c) has a voltage floor of $100\mathrm{nV}$ imposed
while (d) has one at the $10 \mu \mathrm{V}$ level.}
\label{fig:collapse_above}
\end{figure}

Fig.\ \ref{fig:collapse_above}(a) shows the results of scaling for
$T_{KT}^{M}=21K$. The collapse is much worse than those presented
in Fig.\ \ref{fig:collapse_comparison}. As the modified
$T_{KT}^{M}$ is increased the data collapse becomes even worse, as
is seen in Fig.\ \ref{fig:collapse_above}(b) for
$T_{KT}^{M}=\mathrm{24K}$. The deviations shown in Figs.\
\ref{fig:collapse_above}(a) and \ref{fig:collapse_above}(b) cannot
be attributed to driving the sample normal since these departures
all occur for the lowest currents. These data clearly do not scale
for higher modified $T_{KT}^{M}$.


However, when we remove the lowest voltage data shown in Fig.\
\ref{fig:IV}, to mimic a less sensitive experiment, and repeat the
scaling analysis, we find that the highest transition temperature
for which collapse occurs increases. This is shown in Figs.\
\ref{fig:collapse_above}(c) and \ref{fig:collapse_above}(d) where
the measurement floors are set to $100 \mathrm{nV}$ and $10 \mu
\mathrm{V}$, respectively.  For the $100 \mathrm{nV}$ cutoff, the
highest $T_{KT}^{M}$ which gives good collapse is about 21K while
for the $10\mu \mathrm{V}$ cutoff it is raised up to 24K.
Measurements with even worse sensitivity would allow us to define
still higher transition temperatures.

\subsection{Implications of further analysis}

The preceding discussion in this section shows that the modified
scaling analysis alone is insufficiently restrictive.  Any trial
value of $T_{KT}^{M}$ below 17.6K yields a good scaling collapse,
and higher $T_{KT}^{M}$ would also work if the voltmeter were less
sensitive.  This has important implications because it could
easily account for the apparent universal result of $z \approx 6$
that Pierson \emph{et al.}~\cite{pierson99_1prb} and Ammirata
\emph{et al.}~\cite{ammirata99_1pc} have determined on various
systems, and in which there has been much recent
interest.~\cite{colonna-romano00_1condmat,medvedyeva00_1prb,medvedyeva01_1physicaC,holzer01_1prb,pierson00_1prb}

These implications can be understood if we consider a comparison
of several hypothetical measurements of the same sample made with
varying sensitivities.  If one would attempt to attribute a low
universal value of $z$ to all these measurements it would be found
that the less sensitive ones would show good agreement with
scaling (like those of Figs.\ \ref{fig:collapse_comparison}(a-d),
\ref{fig:collapse_above}(c), and \ref{fig:collapse_above}(d))
while the more sensitive measurements would not (like those of
Figs.\ \ref{fig:collapse_above}(a), and
\ref{fig:collapse_above}(b)). However, as the value of $z$ is
increased, one would arrive at values that could collapse all the
measurements and universality could be claimed amongst the various
measurements. If various 2D-superconducting systems have the same
flexibility of analysis that we find for the Repaci \emph{et al.}
data (a completely reasonable assumption) then the same arguments
could be applied amongst measurements made on the various systems.
Thus, by choosing a sufficiently large $z$ value, our analysis
implies that one should always be able to obtain universal
agreement amongst different measurements and systems.

Our analysis in this paper investigates in depth the single
measurement of Repaci \emph{et al.}  It would therefore be an
interesting test of our hypothesis to see the results of other
groups rigorously examining the flexibility in analysis of their
data.

Having demonstrated that relying on the modified KTB scaling alone
is clearly unsatisfactory, we now investigate the alternative
finite-size scaling method of Medvedyeva \emph{et
al.},~\cite{medvedyeva00_1prb} Eq.\ (\ref{minn_scaling}); a
scaling method that also applies FFH ideas to the KTB transition.

\subsection{Finite-size scaling}

\begin{figure}
\epsfig{file=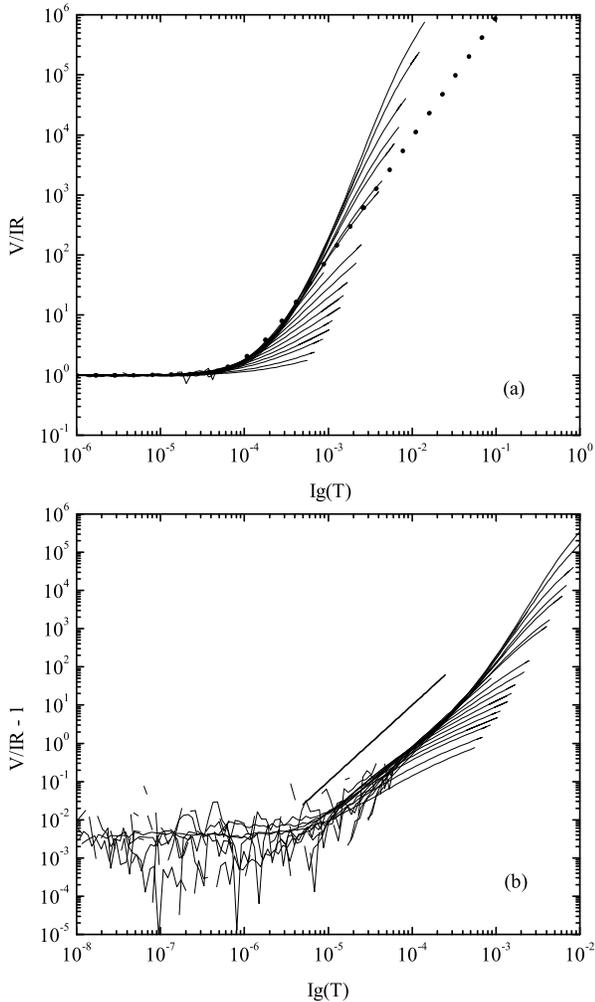,clip=,width=3.1in} \caption{(a) shows the
finite-size scaling collapse of the Repaci \emph{et al.} data with
a function of the form ${V \over {IR(T)}}=1+cI^2$ superimposed.
(b) shows a more sensitive test of a finite-size scaling collapse
with the straight line representing ${V \over {IR(T)}} - 1 \propto
I^2$.  Both (a) and (b) demonstrate that finite-size scaling
agreement with measured data is due to trivial analytic behavior,
like that described by the expansion in Eqs.\
(\ref{V_expand_minnhagen}) and (\ref{V_expand_minnhagen2}).}
\label{fig:finite}
\end{figure}

To examine the data through Medvedyeva \emph{et
al.}'s~\cite{medvedyeva00_1prb} theory we will ignore the size
dependance of the scaling since the measurements are made on the
same sample.  We thus rewrite Eq.\ (\ref{minn_scaling}) as

\begin{equation}
{{V} \over {IR(T)}} = h(I {\mathrm{g}}(T)),
\label{minn_scaling_appl_to_repaci}
\end{equation}
with ${\mathrm{g}}(T)$ unspecified by the theory.  In the work of
Medvedyeva \emph{et al.} they determine that a ${\mathrm{g}}(T)$
given by
\begin{equation}
{\mathrm{g}}(T) \propto R(T)^{-\alpha}, \label{connection_scaling}
\end{equation}
with $\alpha \approx 1/6$ provides a good collapse of the Repaci
data in the low current regime.  We have reproduced this scaling
collapse in Fig.\ \ref{fig:finite}(a), where the solid lines are
Repaci \emph{et al.}'s data at 23.0K and above.

Although this seems to show good agreement with the scaling
theory, there is intrinsic flexibility in this analysis.  To
demonstrate this flexibility, we suppose we have any group of
non-linear $I-V$ isotherms made on the same sample which may or
may not be applicable to the scaling theory of Medvedyeva \emph{et
al.}  The simplest scenario would be that the $I-V$ curves at any
single temperature are analytic and can be expanded in the form

\begin{equation}
V=R(T)I+R_3(T)I^3+R_5(T)I^5+\ldots, \label{V_expand_minnhagen}
\end{equation}
where we have kept only odd terms in the expansion since the
simplest behavior is to assume the voltage changes sign upon
reversing the applied current.  (We further point out that the
measurement techniques of Repaci \emph{et al.} assume this
antisymmetry.)

We can recast Eq.\ (\ref{V_expand_minnhagen}) into the form of
Eq.\ (\ref{minn_scaling_appl_to_repaci}) by dividing by $R(T)I$,
which yields
\begin{equation}
{V \over {IR(T)}}=1+\left( I \sqrt{{{R_3(T)} \over {R(T)}}}
\right)^2+ \left( {{R_5(T)} \over {R(T)}} \right) I^4+\ldots
\label{V_expand_minnhagen2}
\end{equation}
The above relation will always satisfy the Medvedyeva \emph{et
al.} scaling requirements at low currents up to the second term in
the expansion by identifying $\sqrt{{{R_3(T)} \over {R(T)}}}$ as
${\mathrm{g}}(T)$.  Since the form of ${\mathrm{g}}(T)$ is
completely unspecified by the scaling, it can always be determined
by $\sqrt{{{R_3(T)} \over {R(T)}}}$ and thus this finite-size
scaling should always hold for analytic $I-V$ behavior.

To test whether this is the sort of trivial agreement being seen
in the data collapse of Fig.\ \ref{fig:finite}(a), we superimpose
a function of the form ${V \over {IR(T)}}=1+cI^2$, which is simply
the first two terms of the expansion of Eqs.\
(\ref{V_expand_minnhagen}) and (\ref{V_expand_minnhagen2}); with
$c$ a constant fitting parameter that determines the unspecified
${\mathrm{g}}(T)$. Clearly, this function fits the data collapse
well and indicates that the apparent success of this scaling could
be due to the intrinsic flexibility of the analysis.


A better test of the Medvedyeva \emph{et al.} scaling is to
subtract one from ${V \over {IR(T)}}$.  Since the low current
region of the data collapse is expected to be one this should give
a more sensitive examination of the data at small currents,
\emph{i.e.}, the only region where the Medvedyeva \emph{et al.}
finite-size scaling is expected to be valid.  This is plotted in
Fig.\ \ref{fig:finite}(b), where we find that the region over
which the collapse is achieved goes approximately as $I^2$;
\emph{i.e.}, parallel to the solid line. This means that we are
essentially just collapsing the the second (\emph{i.e.},
quadratic) term of the expansion in Eq.\
(\ref{V_expand_minnhagen2}). This is further indication that the
successful scaling is due to trivial analytic behavior as in Eq.\
(\ref{V_expand_minnhagen}) or Eq.\ (\ref{V_expand_minnhagen2}).

Having noted that, we point out that Medvedyeva \emph{et
al.}~\cite{medvedyeva00_1prb} originally proposed that their
scaling is essentially the same as the modified KTB scaling
analysis under certain conditions, most notably that $R(T)$
behaves according to Eq.\ (\ref{connection_scaling}) with $\alpha$
a temperature independent constant equal to $1 / z$.  (This
analogy is of course only valid experimentally in the low current
and high temperature regime due to the need for a low current
linear resistance value in Eq.\ (\ref{minn_scaling}).)  Under
these conditions Medvedyeva \emph{et al.}~\cite{medvedyeva00_1prb}
argue that their finite-size scaling behavior can appear to have a
transition.  They call this a ``ghost" transition because it
should only appear to occur whereas they argue that their
finite-size scaling is actually the correct physical description.

The fact that we find no significant agreement between the
Medvedyeva \emph{et al.} finite-size scaling and the measurements
analyzed here means support for a ``ghost" transition is also
lacking experimentally; since the latter depends on the existence
of the former. That said, we find however that the parameters of
an assumed ``ghost" transition are not flexible when we analyze
the measurements. By using the various scaling tests discussed in
Ref.\ \onlinecite{medvedyeva00_1prb} on the Repaci data, we find
that the value of $1 / \alpha$ in Eq.\ (\ref{connection_scaling})
is well restricted to lie between five and eight.  The fact that
the second term in the Taylor expansion of Eq.\
(\ref{V_expand_minnhagen2}) has this restricted behavior may
indicate some sort of underlying physical basis. Further studies
on experiments with varying sizes might be suitable for
investigating this possibility, as one could utilize the full
finite-size scaling form of Eq.\ (\ref{minn_scaling}), as has been
done for the simulations of Refs.\ \onlinecite{medvedyeva00_1prb}
and \onlinecite{medvedyeva01_1physicaC}.

If the Medvedyeva \emph{et al.} scaling had succeeded in Fig.\
\ref{fig:finite} over regions that the simple analytic behavior
could not have described, then this would have been evidence for
non-trivial agreement with theory.  Thus, this is a sort of
criterion that could be used to determine whether the finite-size
scaling of Medvedyeva \emph{et al.} is non-trivial.

Having motivated a criterion for the finite-size scaling, in the
next section we will address the subtle issue of a criterion for
determining a non-trivial modified KTB transition.  This criterion
will later be used in Sec.\ \ref{sec:discussion} to motivate a
different criterion for determining the existence of a
conventional KTB transition.


\section{An Experimental Criterion for Observing The Modified KTB Transition}
\label{sec:prediction}

In the previous section we showed that data collapses and $R_L$
fits alone are not sufficient to indicate a modified KTB phase
transition. It remains a possibility that such a phase transition
is present, but the scaling analysis does not uniquely determine
it. What is needed is an unambiguous signature for a phase
transition--something to differentiate between true and false
scaling agreements.

Each isotherm in Fig.\ \ref{fig:IV} collapses onto only a small
portion of $\varepsilon_{\pm}$ in Fig.\
\ref{fig:collapse_comparison}(a). In the low current direction of
the collapses the isotherms are cut off by the sensitivity floor
of the experiment. For the region below the sensitivity floor, one
of two possibilities would occur if more sensitive measurements
were made. These measurements would either collapse onto the
scaling function, indicating a real transition, or deviate,
indicating that the original collapse was simply a product of the
experimental resolution limit.

To see what a real transition would look like, we can use the
scaling functions obtained from higher-voltage data to predict
behavior at voltages smaller than the experimental
resolution.\cite{strachan01_prl} We do this by choosing a
temperature $T$, which specifies the appropriate scaling function
to use from a data collapse, like the one in Fig.\
\ref{fig:collapse_comparison}(a). Since $T_{KT}^{M}$ and $b$ are
found from the previous fits, choosing a value of $I$ determines
$I/{\xi T}$, and thus specifies a point on the $x-axis$. This $x$
value and the scaling function determine a $y$ value. The only
unknown in $y$ is $V$, which is thus determined. We emphasize that
if $I$ is large enough, this procedure just returns the measured
value of $V$, but if $I$ is small enough, it gives an extrapolated
small value for $V$ which is beyond the resolution of the
voltmeter. To perform this extrapolation we used a data collapse
with $z=6$, which is the value suggested by Pierson \emph{et
al.}~\cite{ammirata99_1pc,pierson99_1prb} The other two parameters
which collapse the data for this $z$ are $T_{KT}^{M}=18.183
\mathrm{K}$ and $b=5.37$.

\begin{figure}
\epsfig{file=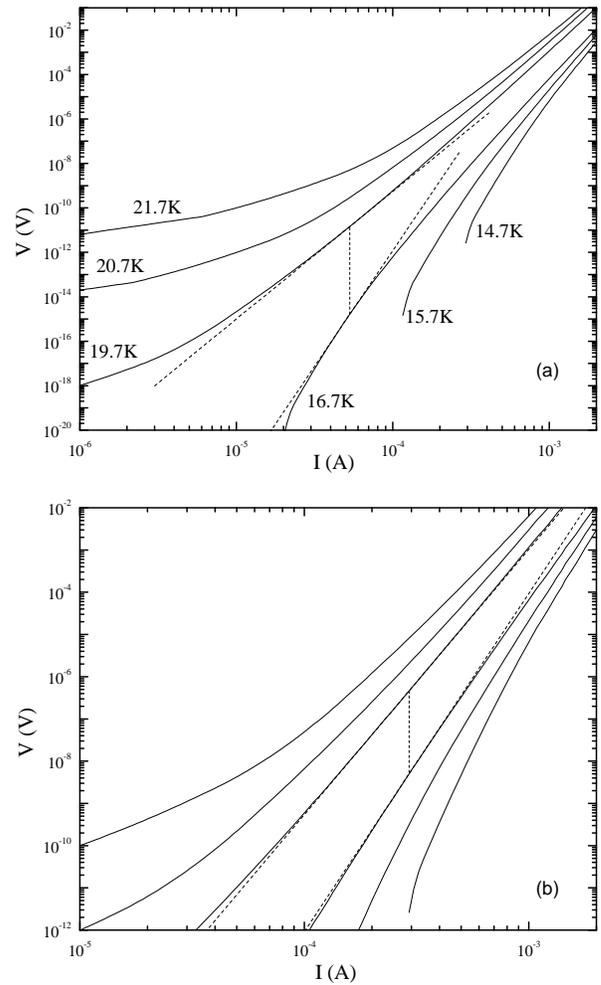,clip=,width=3.1in} \caption{Simulated
\protect\loglog\ isotherms. The temperatures of the isotherms are
marked in (a) with a dashed vertical line drawn between one of the
pairs. The other dashed lines in (a) are tangents drawn to these
isotherms at the location of intersection with the vertical
marker. (b) shows the same curves but with a sensitivity floor
equal to 1pV.} \label{fig:sim_IV}
\end{figure}

The results of the extrapolation are shown in Fig.\
\ref{fig:sim_IV}(a).  Our first observation is that the
extrapolated data from a modified KTB scaling analysis do not show
the $V \sim I^{a(T)}$ behavior expected below $T_{KT}^{M}$ down to
$10^{-20} \mathrm{V}$. Instead, we find that the voltage goes to
zero faster than a power of $I$ as $I \rightarrow 0$, clearly
apparent in the negative concavity for $T < T_{KT}$.  This is
further testament to the incompatibility of conventional KTB
theory and FFH scaling, as outlined earlier in this paper.

Following the arguments of Ref.\ \onlinecite{strachan01_prl}, we
observe next that the extrapolated curves display a signature not
seen in the measured data. Isotherms with equal $\left| (T -
T_{KT}^{M}) / T_{KT}^{M} \right|$ have opposite concavities
\emph{at the same current level}. We demonstrate this in Fig.\
\ref{fig:sim_IV}(a): The dashed vertical line (constant current)
is drawn between isotherms 1.5K on either side of the critical
temperature. The lines tangent to these isotherms clearly show the
opposite concavity. Two other pairs of isotherms are also shown in
Fig.\ \ref{fig:sim_IV}(a) with opposite concavities at the same
current levels.  Fig.\ \ref{fig:sim_IV}(b) clearly demonstrates
that a true data collapse should show this property whether the
voltage resolution is $10^{-20}$ or at $10^{-12}V$. In either case
$T_{KT}^{M}$ would be restricted to lie well within the two inner
curves at 16.7K and 19.7K, and would thus be independent of the
resolution limit of the experiment.

Repaci \emph{et al.}'s data in Fig.\ \ref{fig:IV} has positive
concavity in all the $I-V$ isotherms above 19.5K (and over a 12.5K
range) at approximately $1.5 \times 10^{-4} \mathrm{A}$. For a
transition to exist, this trend must cease since, below
$T_{KT}^{M}$, isotherms are not expected to have ohmic tails. To
show that this trend ceases it is necessary to see negative
concavity for an isotherm below $T_{KT}^{M}$, while one above, at
the same current, and with equal $\left| (T - T_{KT}^{M}) /
T_{KT}^{M} \right|$ has a positive concavity. We require that the
relevant temperature scale of the transition, $\left| (T -
T_{KT}^{M}) / T_{KT}^{M} \right|$, be equal since both isotherms
must be in the critical region, excluding the possibility of
comparing critical to non-critical behavior.


Repaci \emph{et al.}'s data shows no evidence of this necessary
signature about $T_{KT}^{M}$. We conclude that the measurements
are not consistent with scaling.

Having proposed a criterion for determining a modified KTB
transition, we use it in the next section to help motivate a
criterion for determining a conventional KTB transition.


\section{Applicability of analysis to conventional KTB transition.} \label{sec:discussion}

The quintessential characteristic of a KTB transition is the
universal jump in the fully renormalized superfluid density from

\begin{equation}\label{jump}
{{\rho_{s0}} \over (\epsilon_{\infty} / \epsilon_{o})} =
{{m^{*}n^{*}_{s0}} \over (\epsilon_{\infty} /
\epsilon_{o})}={{2{m^{*}}^{2}k_{B}T_{KT}} \over {\hbar^{2}}}
\end{equation}
at $T^{-}_{KT}$ to zero at $T^{+}_{KT}$ (see for example Ref.\
\onlinecite{young94_1nato}). In the above equation, $n^{*}_{s0}$
is the bare Cooper pair density per area while $m^{*}$ is the
Cooper pair mass.

This jump has been reported in many superconducting systems for
both finite frequency and DC $I-V$
measurements.~\cite{norton93_1prb,yeh89_1prb,vadlamannati91_1prb,stamp88_1prb,ying90_1prb,kim89_1prb,park00_1prb,matsuda92_1prl,walkenhorst92_1prl,ammirata99_1pc,ban89_1prb,teshima92_1pc,pierson99_1prb,fiory88_1prl,martin89_1prl,harris91_1prl,mooij94_1nato,garland87_1prb,kadin83_1prb,zant90_1jlt,herbert97_1jltp,herbert98_1prb,higgins00_1prb,resnick81_1prl,epstein81_1prl,hebard83_1prl,hebard80_1prl}
However, the jump is only expected in the limit of $\omega
\rightarrow 0$ and $I \rightarrow 0$.  At finite frequencies and
currents the jump is somewhat rounded which makes determination of
the KTB transition less obvious.  In fact, it has recently been
suggested~\cite{rosario00_1prb} that many reports of the KTB
transition do not show the universal jump or transition at all.

\subsection{Motivating a KTB concavity criterion}

To address this concern, we propose a criterion for determining
whether this jump exists (as opposed to a possible finite size
effect~\cite{repaci96_1prb,herbert98_1prb,rosario00_1prb,holzer01_1prb})
in DC $I-V$ measurements.~\cite{lobb_review_mention}

We motivate a criterion by employing the one developed in the last
section with one change. Since a KTB transition predicts $V
\propto I^{a(T)}$ below $T_{KT}$, we expect that the below
$T_{KT}$ isotherms have \emph{zero} concavity on a $\log - \log$
plot at the same applied currents, while ones above have positive
concavity.~\cite{zero_concavity} We will refer to this criterion
as the ``KTB concavity criterion," to distinguish it from the
``opposite concavity criterion," of the last section and Ref.\
\onlinecite{strachan01_prl}. This KTB concavity criterion is
further supported by a consideration of the important length
scales of the problem, as we will show below in Sec\
\ref{further_support}.

\subsection{What is the need for the KTB criterion?}

First we will consider the case of a finite-size effect at
temperatures much below $T_{KT}$.  In this regime there should not
be a cubic $I-V$ power law because we are far below the real
$T_{KT}$ for an infinite 2D sample.  However, we will now argue
that evidence for a cubic $I-V$ power law will always be found in
this finite size dominated regime despite the fact that a KTB
transition does not exist here.

In this low temperature regime we would expect $V \propto
I^{a(T)}$ at all finite applied currents such that
$\lambda_{\perp} > r_c > \xi_{GL}$,~\cite{min_probe_length} where
$\xi_{GL}$ is the approximate normal core size and also the
smallest separation for a bound pair. $V \propto I^{a(T)}$
because, as $T$ is lowered below $T_{KT}$, renormalization effects
due to fluctuating vortex pairs diminishes quite
rapidly~\cite{sujani94_1prb,davis90_1prb} and so $\epsilon(r_c)
\approx \epsilon_{\infty} \approx \epsilon_0$.

At sufficiently low temperatures, as the current is lowered the
measurement floor is reached before $r_c$ becomes comparable in
size to $\lambda_{\perp}$ or the sample size. Therefore, in this
regime we only expect to find the power-law dependence expected
below $T_{KT}$.  As higher temperature $I-V$ curves are
investigated, but still below the true $T_{KT}$, lower currents
produce measurable voltages.~\cite{lower_currents_measurable}
According to Eq.\ (\ref{rc}), this means that larger length scales
could be probed, \emph{i.e.}, larger $r_c$. When $r_c$ becomes
comparable to $\lambda_{\perp}$ or the sample size we would expect
ohmic tails due to the existence of free vortices. Since these
finite-size induced ohmic tails will emerge smoothly (due to the
statistical nature of unbinding) from isotherms with $a(T) > 3$,
one must necessarily pass an $I-V$ curve which seems to have the
power of 3.   Thus, one would conclude that a rounded jump in
${{\rho_{s0}} \over (\epsilon_{\infty} / \epsilon_{o})}$ is
measured; the sort typical of a true KTB transition, even though
no true transition occurs.

\subsubsection{Illustrating need for criterion with example}

\begin{figure}
\epsfig{file=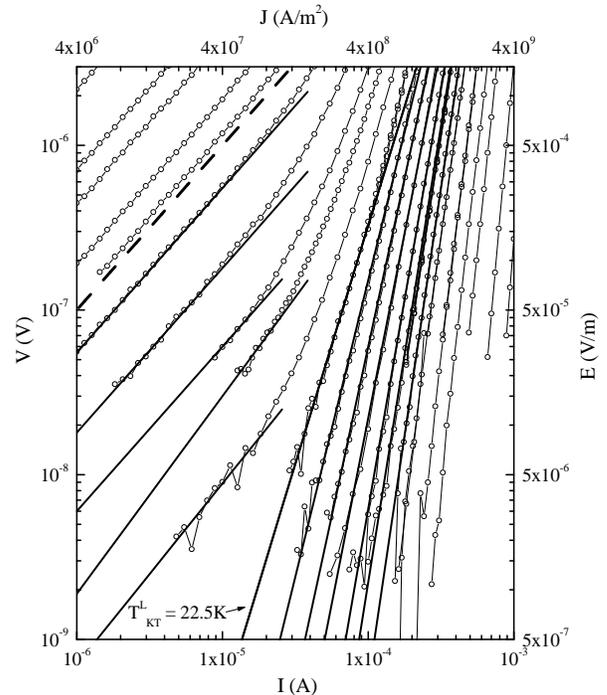,clip=,width=3.1in} \caption{A blow
up of Fig.\ \ref{fig:IV} near where one might conclude that a KTB
transition occurs.  The solid lines are power-law fits to the
isotherms.} \label{KT_conv_anal}
\end{figure}

\begin{figure}
\epsfig{file=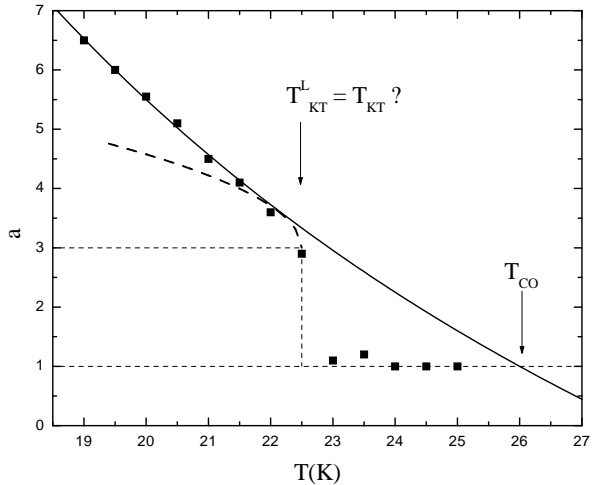,clip=,width=3.1in} \caption{The $a(T)$
derived from Fig.\ \ref{KT_conv_anal}.  The solid line is a mean
field (or bare superfluid density) fit, while the curved dashed
line schematically represents the square root cusp expected just
below $T_{KT}$.~\cite{kadin83_1prb}} \label{a(T)}
\end{figure}

This is illustrated in Fig.\ \ref{fig:IV} by fitting the power of
2.9 (\emph{i.e.}, approximately 3) to the low-current regime of
the curve at 22.5K ($T_{KT}^{L}$). In Fig.\ \ref{KT_conv_anal} we
show a blow up of Fig.\ \ref{fig:IV} with power law fits to the
low current portions of the $I-V$ curves.  Over three decades in
voltage we \emph{seem} to have very good power-law fits, as
expected for the KTB transition. We draw further connection with
the conventional KTB analysis by plotting in Fig.\ \ref{a(T)} the
$a(T)$ derived from these power-law fits.  Remarkably, both plots
seem to indicate, in the usual way, that a universal jump in the
superconducting electron density occurs.


This is a significant observation in that the temperature where we
have just determined a conventional KTB transition to lie (22.5K)
is midway between 17.6K ($T_{KT}^{M}$) where Pierson \emph{et
al.}~\cite{pierson99_1prb} and Ammirata \emph{et
al.}~\cite{ammirata99_1pc} argue for a modified KTB transition,
and 27K ($T_{KT}^{H}$) where Medvedyeva \emph{et
al.},~\cite{medvedyeva00_1prb} argue that a conventional KTB
transition would have existed had finite-size effects not obscured
it. Furthermore, according to Repaci \emph{et al.} and Medvedyeva
\emph{et al.} the regime we focus on in Fig.\ \ref{KT_conv_anal}
is dominated by finite size effects and should not support a
conventional KTB transition.

Thus, despite the fact that various scenarios have been proposed
that are incompatible with a KTB transition at 22.5K, we do in
fact find very good agreement with theory for one.  In light of
this perplexing situation we ask if there are other criterions
that we could use to determine whether our agreement with KTB
theory is in fact valid?

\subsubsection{How about other criterions?}

To start with, the universal jump condition shown in Fig.\
\ref{a(T)} is clearly not sufficient as we have just seen. Another
possibility is the fit of the linear resistive tails; however, we
have already explained in Sec.\ \ref{RL_fit_section} that these
fits are far too flexible.   A third possibility is the square
root cusp of the super-electron density just below the
transition.~\cite{kadin83_1prb}  However, this square root cusp is
typically very difficult to determine experimentally (see for
example the fits in Fig. 2 of Ref.\ \onlinecite{epstein81_1prl}).
This may be due to the functional form of the super-electron
density being highly dependant on the experimental length scale
probed.  For example, the recent theoretical work of Pierson
\emph{et al.}~\cite{pierson00_1prb} and the analysis of Kaden
\emph{et al.}~\cite{kadin83_1prb} show a stark change in the shape
of the super-electron density near $T_{KT}$ for different
characteristic length scales, while the super-electron density at
$T_{KT}$ remains relatively unchanged.

Considering this lack of an adequate criterion, we will now give
further justification for our KTB concavity criterion. Since two
of the three conflicting interpretations of the Repaci \emph{et
al.} data claim that the regime of Fig.\ \ref{KT_conv_anal} is
dominated by finite-size effects, we will focus on differentiating
between size effects and true KTB behavior.

\subsection{A further argument in support of the KTB concavity
criterion} \label{further_support}

Since finite-size effects should dominate when $r_c$ is roughly of
size $\lambda_{\perp}$, we can substitute $\lambda_{\perp}$ into
Eq.\ (\ref{rc}) and solve for the current where we cross over to
these effects, \emph{i.e.}
\begin{equation}\label{Ico_finite_size}
j_{co}^{\lambda_{\perp}} \approx {\Phi_0 \over {2 \pi \mu_0
\lambda_{\perp}^2}}.
\end{equation}
Typically we should expect $\lambda_{\perp}$ to be only weakly
temperature dependent in this regime and, thus, we should also
expect $j_{co}^{\lambda_{\perp}}$ to be roughly constant as a
function of temperature.  In the case where $\lambda_{\perp}$ is
temperature dependent, $\lambda_{\perp}$ should grow as
temperature increases. Thus, we conclude that
$j_{co}^{\lambda_{\perp}}$ will decrease or remain constant as $T$
increases for finite-size induced unbinding.

For a true KTB transition, where the ohmic tails are due to a
finite correlation length when $T > T_{KT}$, $j_{co}^{KT}$ should
have the completely opposite temperature dependence, \emph{i.e.},
increasing as $T$ grows. A rough argument supporting this goes as
follows:

At temperatures sufficiently close to and above $T_{KT}$ we expect
changes to occur at large length scales determined by the
diverging $\xi_+$. That is, the largest pairs are the first to
unbind in going up through $T_{KT}$.  Thus, when probing at
sufficiently small scales, \emph{i.e.}, with sufficiently large
$j_{2D}$ and for $T \gtrsim T_{KT}$, we should not expect large
deviations from the $E \sim j_{2D}^{3}$ behavior expected at
$T_{KT}$. We approximate this behavior near $T_{KT}$ for large
driving currents as
\begin{equation}\label{Vnear_Tkt}
  E \approx C_{Ej}j_{2D}^{3},
\end{equation}
where $C_{Ej}$ is a constant and $E$ is the electric field.
However, we also expect the temperature dependence of the ohmic
tails to vary as
\begin{equation}\label{Rnear_Tkt}
{{E} \over {j_{2D}}} = \rho \approx {\rho_0 \over \xi_+^2}= \rho_0
e^{-2\sqrt{bT_{KT}/(T-T_{KT})}},
\end{equation}
according to Eq.\ (\ref{eq:R_L}) with $z=2$. Eliminating $E$
between Eqs.\ (\ref{Vnear_Tkt}) and (\ref{Rnear_Tkt}), we obtain a
rough estimate for the crossover current between power law and
ohmic behavior very close to $T_{KT}$, \emph{i.e.},
\begin{equation}\label{Ico}
  j_{co}^{KT} \approx \sqrt{\rho_0 \over C_{Ej}}e^{-\sqrt{bT_{KT}/(T-T_{KT})}}.
\end{equation}
This equation says that this crossover should occur at increasing
values of current as the temperature is raised above $T_{KT}$. In
the crossover regime between power law and ohmic $\log V - \log I$
curves we expect positive concavity.  Since the extent of this
regime will increase with temperature there should be some
isotherms with $T<T_{KT}$ which have zero concavity at the same
applied current. This is the extension of the concavity criterion
applied to a true KTB transition discussed earlier in this
section.

On the other hand, the crossover and its associated positive
concavity due to finite size effects would occur at either a
constant applied current or a decreasing current as temperature is
raised above $T_{KT}$.  This implies that we would \emph{not} see
positive concavity for higher temperature isotherms while lower
ones maintain zero concavity.  The occurrence of either behavior
can be used to determine the cause of the ohmic tails. Again,
since the relevant temperature scale is $\left| T-T_{KT} \over
T_{KT} \right|$ we should be comparing isotherms above and below
$T_{KT}$ with this value equal, not to confuse critical with
non-critical behavior.

Despite our rough arguments, it is instructive to take another
look at the $I_{co}$ plotted on Fig.\ \ref{fig:IV}.  At low
temperatures we see that the crossover current is roughly constant
as a function of temperature.  According to our arguments this
would indicate that this results from a finite size effect, in
agreement with Repaci \emph{et al.}'s conclusion. Furthermore, as
temperature is raised above about 32K, we begin to clearly see
that this crossover current increases.  Interestingly, the
increase in $I_{co}$ sets in roughly where the high current regime
of the $I-V$ curves has a power of approximately 3, \emph{i.e.}, $
T \approx T_{KT}^{H} \approx 27.5K$.  These observations give
further support to Repaci \emph{et al.}'s~\cite{repaci96_1prb} and
Medvedyeva \emph{et al.}'s~\cite{medvedyeva00_1prb} view that
finite-size effects obstruct the KTB transition from occurring.

\subsection{Putting the KTB concavity criterion to the test}

As a test of our criterion we review a theoretical work and three
simulations which report $I-V$ isotherms near a KTB transition.
First we consider the renormalization-group analysis of current
induced vortex pair unbinding by Sujani \emph{et
al.}~\cite{sujani94_1prb} The results of their analysis, which is
based on a model without finite-size effects included, clearly
demonstrates a positive concavity for higher temperature isotherms
(see Fig.~5(b) in Ref.\ \onlinecite{sujani94_1prb}), while
isotherms below $T_{KT}$ and with equal $ \left| T-T_{KT} \over
T_{KT} \right|$ do not show concavity.

We next consider the simulations of $I-V$ curves by Holzer
\emph{et al.} which contain both a true KTB transition plus finite
size effects.~\cite{holzer01_1prb}  At higher currents their $I-V$
curves are dominated by characteristics of a true KTB transition.
In this regime we see that our KTB concavity criterion is clearly
satisfied.  However, at lower currents their $I-V$ characteristics
are dominated by size effects and the concavity criterion is no
longer satisfied.

A third test is the numerical simulations of Colonna-Romano
\emph{et al.}~\cite{colonna-romano00_1condmat} Like the $I-V$
curves of Holzer \emph{et al.}, they too show true KTB behavior at
high currents and finite-size effects at lower currents. Again, in
the KTB regime our criterion is satisfied while in the finite-size
dominated region it is not.

Our final check uses the $I-V$ simulations of Medvedyeva~\emph{et
al.}~\cite{medvedyeva00_1prb}  Like the simulations of
Colonna-Romano \emph{et al.} and Holzer \emph{et al.}, these
demonstrate finite-size effects at low currents which do not
satisfy the KTB concavity criterion. At high currents, there is a
slight opposite concavity which may be due to the saturation of
pair unbinding.~\cite{colonna-romano00_1condmat} That is, all the
pairs tend to be unbound at high currents, which should cause
$I-V$s to become ohmic in this regime and thus have negative
concavity.

Based on these four works, it seems that evidence of a KTB
transition is found when the KTB concavity criterion is satisfied.
Having motivated the KTB concavity criterion and found support for
it from theoretical studies and simulated $I-V$ curves, we now
review the literature to see if it is satisfied experimentally.

\subsubsection{Testing experimental results with criterion}

First we review those reports of a KTB transition in conventional
2D superconductors and \jj\ arrays.  In this case we find that the
criterion is clearly satisfied for the superconductors in Refs.\
\onlinecite{hebard83_1prl} and \onlinecite{epstein81_1prl} and for
the arrays in Ref.\ \onlinecite{resnick81_1prl}, while for many
other reports the criterion is not satisfied. Thus, our criterion
is consistent with a KTB transition in at least some conventional
type II superconductors and arrays.

Having shown that the KTB concavity criterion can be
experimentally satisfied, we now address the cuprate
superconductors.

Whether a KTB transition exists in 2D cuprates has been an
unresolved issue (see discussion in Ref.\
\onlinecite{mooij94_1nato}).  Furthermore, many papers report
$I-V$ critical isotherms with $a(T_{KT})$ greater than 6, which is
essentially the basis for Pierson \emph{et al.}'s
proposal~\cite{ammirata99_1pc,pierson99_1prb} that $z
> 5$. However, we find only one report of $I-V$ measurements of a
2D cuprate superconductor, Ref.\ \onlinecite{matsuda92_1prl},
which satisfies our criterion.

\begin{figure}
\epsfig{file=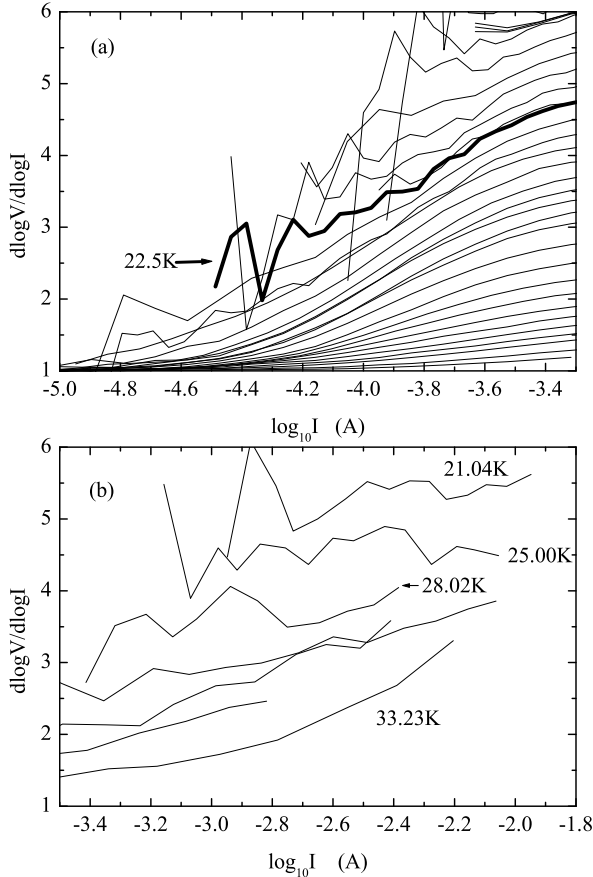,clip=,width=3.1in} \caption{(a)
are the derivatives of the $I-V$ curves in the regime examined in
Fig.\ \ref{KT_conv_anal}.  (b) are the
derivatives~\cite{obtaining_matsuda_data} from $I-V$ measurements
of Ref.\ \protect\onlinecite{matsuda92_1prl}.}
\label{fig:compare_data_sets}
\end{figure}

To demonstrate the difference between the data of Ref.\
\onlinecite{matsuda92_1prl} and the those of Repaci \emph{et al.}
we plot ${d \log V} / {d \log I}$ vs
current~\cite{obtaining_matsuda_data} for both sets in Fig.\
\ref{fig:compare_data_sets}.  In Fig.\
\ref{fig:compare_data_sets}(a) we make this plot for the $I-V$
regime of Repaci \emph{et al.}'s data shown in Fig.\
\ref{KT_conv_anal}. Despite the fact that the typical KTB analysis
shown in Fig.\ \ref{a(T)} indicates there is a transition, these
data fail our KTB concavity criterion.  This is evident from Fig.\
\ref{fig:compare_data_sets}(a) where all the ${d \log V} / {d \log
I}$ have positive slope (\emph{i.e.}, positive concavity for $I-V$
curves) both above and below the $T_{KT}$ determined through
Figs.\ \ref{KT_conv_anal} and \ref{a(T)}.  This is in contrast to
a similar plot for the data of Ref.\ \onlinecite{matsuda92_1prl}
in Fig.\ \ref{fig:compare_data_sets}(b).  Here we see that the
low-temperature ${d \log V} / {d \log I}$ plots level off between
three and four; a value in reasonable accord with the conventional
KTB theory.~\cite{halperin79_1jltp}  This could explain the poor
fit of this data to the modified KTB scaling analysis (see Ref.\
\onlinecite{pierson99_1prb}) which had been analyzed using the
unusually large power at the critical isotherm of approximately 7,
as opposed to 3.


We also emphasize that the axes of Figs.
\ref{fig:compare_data_sets}(a) and \ref{fig:compare_data_sets}(b)
have the same span; ${d \log V} / {d \log I}$ goes from one to six
while $\log_{10} I$ spans 1.8. The only difference in the two
plots is that the range of Matsuda \emph{et al.}'s data (Ref.\
\onlinecite{pierson99_1prb}) is centered at higher applied
currents. Since one generally expects the length scale probed to
be smaller for larger applied current densities, this may may
simply indicate that the data of Matsuda \emph{et al.} is in a
regime where finite-size effects are not probed.


\section{Summary}
\label{sec:summary}

We find that the modified KTB scaling
analysis~\cite{ammirata99_1pc,pierson99_1prb} gives inconclusive
results when applied to measurements of a $\mathrm{12 \AA}$ thick
\ybco\ film. The choice of a modified critical temperature and
exponents is arbitrary to within factors of 3 or more and the
results depend significantly on experimental sensitivities.  We
argue that this flexibility in analysis could be the source for
the large and apparently universal $z$ exponent values.

We propose a criterion necessary to ensure that a scaling analysis
is not afflicted with the same problems as the modified KTB
scaling analysis. We use this to motivate a different criterion
for determining a conventional KTB transition. We find that some
two-dimensional superconductors satisfy our criterion; however,
many, in particular most of those in the cuprates, do not.


\acknowledgements

We thank S. M. Anlage, A. Biswas, R. L. Greene, H.-J. Kwon, P.
Minnhagen, M. C. Sullivan, and Huizhong Xu for useful discussions
on this work. We also acknowledge the support of the National
Science Foundation through Grant Nos. DMR-9732800 and DMR-9801825.





\begin{references}
\bibitem{standard_KTB_theory}
For the standard KTB theory, $z=2$ and thus $V \propto I^3$ at
$T_c$.
\bibitem{zero_concavity}
There is a technical issue about determining zero concavitity as
opposed to
  positive concavity. Since any curve with positive concavity will, over a
  short enough length of line, seem to have zero concavity, it is important
  that the same range of current used to determine positive concavity for a
  high temperature isotherm is also used to determine a zero concavity isotherm
  below $T_{KT}$.
\bibitem{min_probe_length}
To probe length scales smaller than $\xi_{GL}$ would no longer
investigate the
  inter-vortex interactions which are desired.
\bibitem{lower_currents_measurable}
This assumes that at fixed DC current the dissipation (and thus
voltage) is
  greater as temperature increases. This reasonable assumption is supported by
  the $I-V$ curves in Fig.\ \ref{fig:IV}. We can also see from this data that
  the higher temperature isotherms intersect with the $10^{-9}$V resolution
  limit at lower currents.
\end{references}

%
%

%
%

\end{document}